# МОДЕЛИРОВАНИЕ ДИИЗОПРОПИЛОВОГО ЭФИРА МЕТОДОМ МОЛЕКУЛЯРНОЙ ДИНАМИКИ В РАЗЛИЧНЫХ МЕЖАТОМНЫХ ПОТЕНЦИАЛАХ

2022 г. О.В. Кашурин[a, *], Н. Д. Кондратюк[a, b, c], А. В. Ланкин[b, a], Г. Э. Норман[b, a, c]

[a]Московский физико-технический институт (национальный исследовательский университет), Московская область, Долгопрудный, Россия

[b]Объединенный институт высоких температур РАН, Москва, Россия

[c]Национальный исследовательский университет "Высшая школа экономики", Москва, Россия

*e-mail: kashurin.ov@phystech.edu

Для диизопропилового эфира методом классической молекулярной динамики с использованием трех потенциалов проводится сравнительная оценка точности определения плотности и вязкости. Также исследуется точность определения коэффициентов вязкости при использовании равновесных и неравновесного методов расчета.

Ключевые слова: диизопропиловый эфир, молекулярная динамика, диффузия, вязкость.

## 1. ВВЕДЕНИЕ

Диизопропиловый эфир – простой алифатический эфир с формулой $C_6H_{14}O$ представляет собой бесцветную жидкость, растворимую в органических растворителях. Он используется как органический растворитель в химической технологии при производстве красителей и смол, а также для выделения и очистки различных веществ методом экстракции [1-3]. Добавление диизопропилового эфира к бензину, аналогично добавлению метил-трет-бутилового эфира, позволяет сократить выбросы частиц сажи и моноксида углерода при работе двигателей внутреннего сгорания [4-7].

Диизопропиловый эфир представляет практический интерес не только как промышленный растворитель и добавка к топливу, но и как компонент электрохимических источников тока и систем выделения редких металлов из растворов. Простые эфиры могут стать ключевым компонентом ион-селективных барьеров, представляющих собой "жидкую мембрану" [8], где слой несмешивающейся с водой органической жидкости разделяет два слоя водного раствора. Селективная проницаемость такого слоя по отношению к разным ионам может использоваться как для выделения из водных растворов редких элементов, таких как литий или рубидий, так и при создании новых перспективных электрохимических источников тока типа проточных Red/Ox элементов [9, 10].

Одними из наиболее интересных компонент, которые могут обеспечивать селективность "жидкой мембраны", являются простые эфиры: краун-эфиры и криптанды. Причем, если по диизопропиловому эфиру, в литературе имеется достаточно широкий набор экспериментальных данных, то для краун-эфиров и криптандов они значительно более скудны. Это делает диизопропиловый эфир, удобным объектом отработки методов молекулярно-динамического моделирования, которые в перспективе должны быть пригодны для более широкого класса простых эфиров, включая краун-эфиры и криптанды.

При описании работы ион-селективного барьера типа жидкой мембраны наибольшее значение имеет описание подвижности процессов в среде простого эфира и содержащих его растворов. Причем качество описания процессов переноса хорошо коррелирует с точностью описания транспортных свойств системы: вязкостью и диффузией.

Хотя диизопропиловый эфир является широко используемым и легкодоступным веществом, исследование свойств его смесей с другими органическими веществами всё ещё является актуальной задачей [6, 7, 11]. Многообразие смесей органических веществ, включающих диизопропиловый эфир, делает целесообразным для их исследования использование методов компьютерной химии, таких как метод молекулярной динамики (МД), которые широко используются для исследования структуры и свойств жидкостей [12-15]. Примеры расчета динамических свойств водных растворов приведены в работах [12, 16–18].

Физической основой методов молекулярной динамики являются межатомные потенциалы — математические функции для расчета потенциальной энергии системы атомов с заданными положениями в пространстве. Поэтому точность моделирования при использовании МД определяется тем, насколько верно используемые межатомные потенциалы воспроизводят силовое взаимодействие между атомами и молекулами исследуемого вещества. Верификация широкого многообразия межатомных потенциалов на основе экспериментальных или квантово-химических данных – важная задача современной МД [19–22].

Нужно отметить, что работы, посвященные молекулярно-динамическому моделированию диизопропилового эфира и его смесей с другими веществами, не многочисленны [23, 24]. В большинстве из них авторы проводили исследование только с использованием одного силового поля. Таким образом актуальность данной работы заключается в оценке применимости трех классических потенциалов GAFF (AMBER), OPLS-AA и CHARMM с точки зрения точности воспроизведения методами молекулярной динамики транспортных свойств диизопропилового эфира в широком диапазоне температур и давлений.

В разделе 2 данной работы рассмотрены особенности используемых потенциалов межатомного взаимодействия (п. 2.1), приведены основные параметры и характеристики молекулярно-динамического моделирования (п. 2.2) и описаны используемые методы расчета вязкости (п. 2.3). В разделе 3 представлены результаты оценки плотности (п.п. 3.1) и сжимаемости (п.п. 3.2) диизопропилового эфира в зависимости от давления и температуры (п.п. 3.3), рассмотрена точность определения коэффициентов вязкости при использовании различных методов их оценки (п. 3.4), выполнена сравнительная оценка точности определения транспортных свойств диизопропилового эфира при использовании трех классических потенциалов (п. 3.5).

## 2. МЕТОДЫ МОДЕЛИРОВАНИЯ
### 2.1. Потенциалы межатомного взаимодействия

Для моделирования диизопропилового эфира в работе рассматривались на применимость три межатомных потенциала: General Assisted Model Building with Energy Refinement Force Field (GAFF) [25], Optimized Potentials for Liquid Simulations All-Atom (OPLS-AA) [26] с коррекцией зарядов и Chemistry at HARvard Macromolecular Mechanics (CHARMM) [27] - версия 36.

Параметризация взаимодействий в потенциале GAFF была сформирована с помощью программы Antechamber. Расчет парциальных зарядов на атомах проводился согласно [28], параметризация взаимодействий – согласно [29, 30]. Параметры для OPLS-AA/CM1A сгенерированы на сервере LigParGen [31]. Для парциальных зарядов использовалась коррекция 1.14*CM1A [32, 33]. Константы взаимодействий в CHARMM36 сформированы в CHARM-GUI [34-36].

Все три потенциала включают ковалентные и невалентные взаимодействия между атомами:

$$E = E_{valent} + E_{non-valent}.$$

У GAFF и OPLS-AA/CM1A ковалентные взаимодействия описываются гармоническими колебаниями ковалентных связей, углов между тремя атомами и торсионными взаимодействиями:

$$E_{valent} = E_{bond} + E_{angle} + E_{dihedral}.$$

Данные слагаемые определяются следующим образом:

$$E_{bond} = \sum_{bonds} k_b(l - l_0)^2$$

$$E_{angle} = \sum_{angles} k_a(\theta - \theta_0)^2$$

$$E_{dihedral} = \sum_{dihedrals} \sum_{n}^{N} \frac{k_{d,n}}{2}(1 + \cos(n\varphi - \gamma_n)),$$

где $k_b, k_a, k_{d,n}$ – энергетические константы, $l_0, \theta_0$ - равновесные значения длины связи и угла между атомами соответственно, $\varphi$ - торсионный угол между соответствующими плоскостями, проведенными через тройки атомов.

У потенциала CHARMM36 формула для энергии ковалентных взаимодействий $E_{valent}$ также содержит слагаемые $E_{bond}$, $E_{angle}$ и $E_{dihedral}$, которые описываются теми же функциональными зависимостями. Кроме того, в ней содержатся еще 2 слагаемых - $E_{UB}$ и $E_{improper}$. Слагаемое $E_{UB}$ называется суммой Юри-Брэдли. В ней суммирование идет по всем цепочкам из трех связанных атомов A-B-C. $E_{improper}$ описывает энергию изгиба атомов вне плоскости, а суммирование идет по набору из четырех атомов, которые не связаны последовательно. Эти слагаемые имеют следующий вид:

$$E_{UB} = \sum_{Uray-Bradley} k_u(s - s_0)^2$$

$$E_{improper} = \sum_{impropers} k_{imp}(\psi - \psi_0)^2,$$

где $k_u, k_{imp}$ – энергетические константы, $\psi - \psi_0$ – двугранный угол, образованный четырьмя не связанными последовательно атомами, $s$ – расстояние между атомами A-C, в цепочке атомов A-B-C.

Для всех трех потенциалов невалентные взаимодействия атомов описываются Леннард-Джонсовским и электростатическим потенциалами:

$$E_{non-valent} = \sum_{pairs\ i,j} 4\varepsilon\left(\left(\frac{\sigma}{r_{ij}}\right)^{12} - \left(\frac{\sigma}{r_{ij}}\right)^{6}\right) + k\frac{q_i q_j}{r_{ij}},$$

где $r_{ij}$ – расстояние между взаимодействующими атомами, $\varepsilon$ и $\sigma$ – параметры Леннарда-Джонса, $q_i$ и $q_j$ – заряды атомов.

Для потенциалов GAFF и CHARMM36 при определении параметров $\varepsilon$ и $\sigma$ используется комбинационное правило Лоренца-Бертло [37,38], а для потенциала OPLS-AA/CM1A – среднегеометрическое комбинационное правило [39, 40].

Для атомов внутри одной молекулы, разделенных одной или двумя связями, невалентные взаимодействия не учитываются. По-особому учитываются невалентные взаимодействия для атомов разделённых тремя связями: в OPLS-AA/CM1A для этих атомов взаимодействие Леннарда-Джонса и электростатическое взаимодействие масштабируются с коэффициентом 0.5; в GAFF электростатические взаимодействия и взаимодействия Леннарда-Джонса масштабируются с коэффициентами 0,833 и 0,5 соответственно; в CHARMM36 электростатическое взаимодействие не масштабируется, а взаимодействие Леннарда-Джонса определяется на основе специального набора параметров.

## 2.2. Методы молекулярной динамики

Для проведения моделирования использовался программный пакет GROMACS [41]. Моделирование проводилось для ячейки с 3375 молекулами. Начальная конфигурация задавалось в виде кубической решетки размером $15 \times 15 \times 15$ молекул. Затем проводилось сжатие ячейки до плотности, соответствующей экспериментальной для диизопропилового эфира при заданных температуре и давлении. После сжатия проводился процесс вывода системы на нужную температуру в NVT ансамбле, а затем вывод на нужное давление в NPT ансамбле. Оба процесса моделировались на интервале 200 пс.

Для поддержания NVT ансамбля использовался модифицированный термостат Берендсена [42], а для NPT ансамбля - баростат Берендсена [43]. После установления равновесия в NPT ансамбле рассчитывалась равновесная плотность. Во всех расчетах шаг интегрирования по времени составлял 1 фс. Для исключения краевых эффектов использовались периодические граничные условия.

При моделировании производилась обрезка потенциалов Леннарда-Джонса и электростатического взаимодействия. При этом учитывались поправки к энергии и давлению компенсирующие обрезку потенциалов [44]. Для оценки влияния величины радиуса обрезки на результаты моделирования, расчеты были выполнены при двух значениях этого параметра: 1 нм и 1,2 нм. Было установлено, что при увеличенном радиусе обрезки точность моделирования динамических и других свойств диизопропилового эфира не повышается, поэтому в работе приведены результаты только для 1 нм.

Дальнодействующая часть кулоновского потенциала рассчитывалась методом Эвальда (SPME) [45].

### 2.3. Методы расчета вязкости в МД

В качестве основного метода расчета, использовался метод Грина-Кубо [46, 47]: значение вязкости определяется из соотношения

$$\eta = \frac{V}{kT} \int_0^{+\infty} C_\sigma(t) dt$$

где $V$ – объем расчетной ячейки, $k$ – постоянная Больцмана, $T$ – абсолютная температура системы, $C_\sigma$ – автокорреляционная функция недиагональных элементов тензора напряжений

$$C_\sigma(t) = \langle \sigma_{\alpha\beta}(0) \sigma_{\alpha\beta}(t) \rangle.$$

Угловые скобки означают усреднение по ансамблю и трем взаимно-перпендикулярным плоскостям $xOy, xOz, yOz$. Тензор напряжений рассчитывается по формуле:

$$\sigma_{\alpha\beta} = \frac{1}{V}\left(\sum_{i=1}^{N} m_i v_{i,\alpha} v_{i,\beta} + \sum_{i=1}^{N'} r_{i,\alpha} f_{i,\beta}\right),$$

где $f_{i,\beta}$ – $\beta$ – компонента силы, действующей на $i$-ю частицу, $N$ – количество атомов системы, $N'$ – количество атомов системы и ближайшего образа в случае периодических граничных условий.

Для расчета коэффициентов вязкости этим методом проводилось моделирование в NVT ансамбле в течение 15 нс. Затем из полученной траектории вычислялись 150 статистически независимых автокорреляционных функций $C_\sigma$, которые затем усреднялись, и по полученной функции рассчитывалась вязкость.

Были проверены на применимость два дополнительных метода оценки вязкости: метод Стокса-Эйнштейна и неравновесный метод.

Метод Стокса-Эйнштейна основан на соотношении Эйнштейна:

$$\eta = \frac{kT}{6\pi DR},$$

где $R$ – эффективный радиус молекулы, $D$ – коэффициент самодиффузии, $k$ – постоянная Больцмана, $T$ – абсолютная температура системы. В качестве эффективного радиуса $R$ брался средний радиус гирации молекул.

Коэффициент самодиффузии $D$ вычислялся с помощью формулы Эйнштейна-Смолуховского, описывающей зависимость среднеквадратичного смещения (СКС) молекул на больших временах:

$$\langle(\boldsymbol{r}-\boldsymbol{r_0})^2\rangle = 6Dt,$$

где $\boldsymbol{r}$ – радиус-вектор координаты частицы в момент времени $t$, $\boldsymbol{r_0}$ - радиус-вектор координаты частицы в момент, соответствующий началу отсчета времени.

СКС рассчитывалось в NVT ансамбле в течение 1 нс. При этом из одной траектории вычислялось несколько СКС: начальные точки отсчета брались вдоль всей траектории с периодом 10 пс, для них вычислялось СКС и затем проводилось усреднение.

Неравновесный метод расчета вязкости относится к методам неравновесной молекулярной динамики. Атомы вещества помещаются во внешнее силовое поле вида $A\cos\left(\frac{2\pi z}{L_z}\right)$ [48], а вязкость оценивается как реакция вещества на сдвиговое напряжение. Поскольку внешняя сила совершает работу по смещению атомов нарушается термодинамическое состояние системы. Это может приводить к значительному росту давления [48]. Минимизация этого явления достигается подбором оптимальных значений коэффициента $A$ и размеров расчетной ячейки $L_x, L_y, L_z$. Путем перебора различных параметров было установлено, что для данной системы с потенциалом GAFF оптимальные значения следующие: $A = 0.005$ нм/пс$^2$, $L_x = 5.73$ нм, $L_y = 6.02$ нм, $L_z = 23.4$ нм при 3500 молекулах. При этих параметрах проводился расчет коэффициентов вязкости. Использовался NVT ансамбль, а моделирование проводилось в течение 5 нс.

# 3. РЕЗУЛЬТАТЫ МОДЕЛИРОВАНИЯ

## 3.1. Зависимость плотности диизопропилового эфира от давления

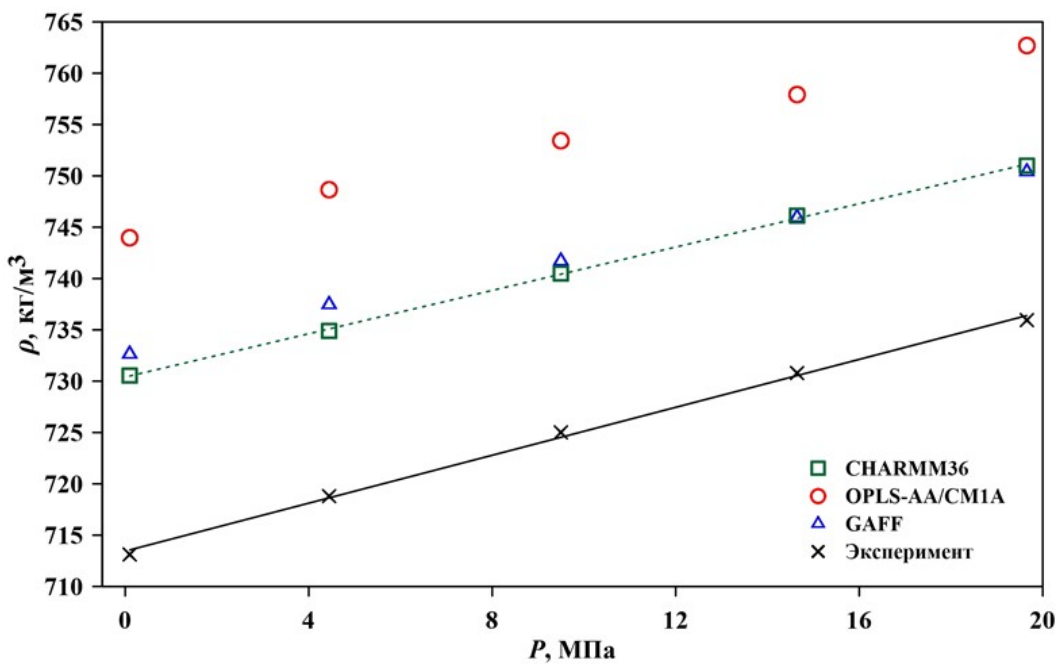

Рис. 1. Зависимость плотности диизопропилового эфира от давления при использовании различных потенциалов. Крестики построены по экспериментальным данным из работы [49]. Прямыми линиями показаны аппроксимации зависимостей по МНК для экспериментальных данных и CHARMM36, который обеспечивает наилучшие среди всех рассмотренных потенциалов значения плотности.

На первом этапе работы была рассчитана плотность при температуре 303.15 К для различных давлений с использованием рассматриваемых потенциалов. По полученным данным построен график зависимости плотности от давления (рис. 1). Экспериментальные значения плотности взяты из [49].

## 3.2. Сжимаемость диизопропилового эфира

Расчет коэффициента сжимаемости $\beta$ диизопропилового эфира в исследуемом диапазоне давлений и температур выполнялся по формуле:

$$\beta = \frac{1}{\rho}\frac{d\rho}{dP},$$

где $\rho$ – плотность системы, $P$ – давление в системе.

Поскольку на исследуемом диапазоне давлений зависимость давления от плотности близка к линейной, то производная $\frac{d\rho}{dP}$ вычислялась как коэффициент наклона аппроксимирующей прямой.

Для оценки коэффициента сжимаемости использовались значения плотности, полученные методом МД с использованием потенциала CHARMM36, поскольку этот потенциал обеспечивал наибольшую точность. Результаты расчетов приведены в Таблице 1.

**Таблица 1.** Значения сжимаемости $\beta$ в зависимости от давления. Коэффициенты $\beta_0$ соответствуют сжимаемости, рассчитанной аналогичным образом, по значениям плотности из [49].

| $P$, МПа | 0.1 | 4.44 | 9.5 | 14.65 | 19.65 |
|---|---|---|---|---|---|
| $\beta, 10^{-6}$ МПа$^{-1}$ | $14.4 \pm 0.4$ | $14.4 \pm 0.4$ | $14.3 \pm 0.4$ | $14.1 \pm 0.4$ | $14.1 \pm 0.4$ |
| $\beta_0, 10^{-6}$ МПа$^{-1}$ | $16.4 \pm 0.4$ | $16.2 \pm 0.4$ | $16.1 \pm 0.4$ | $16.0 \pm 0.4$ | $15.9 \pm 0.4$ |

### 3.3. Зависимость плотности диизопропилового эфира от температуры

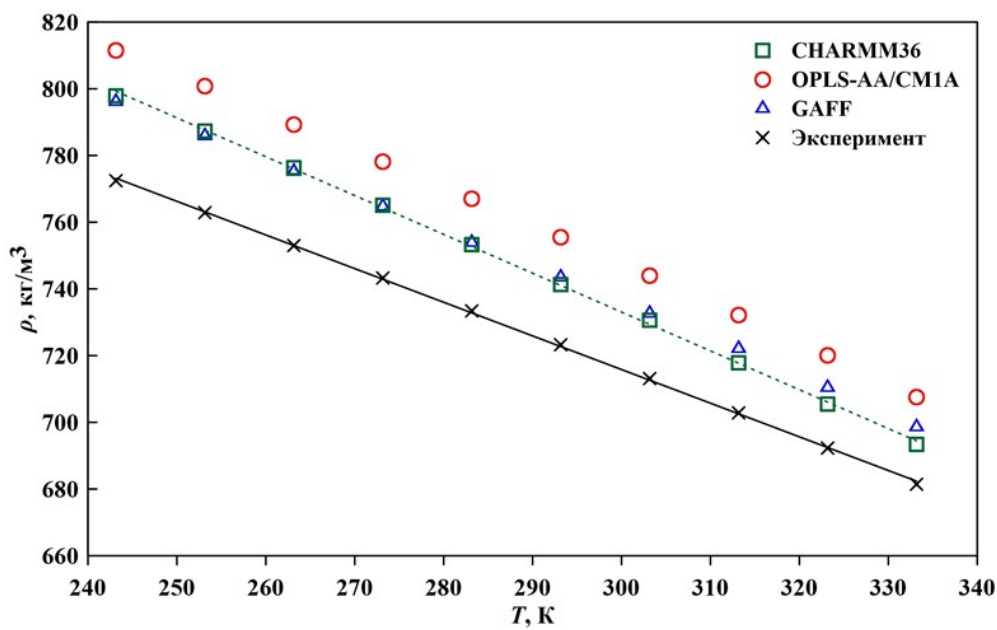

Рис. 2. Зависимость плотности диизопропилового эфира от температуры при использовании различных потенциалов. Крестики построены по экспериментальным данным из работы [49]. Прямыми линиями показаны аппроксимации зависимостей по МНК для экспериментальных данных и CHARMM36, который обеспечивает наилучшие среди всех рассмотренных потенциалов значения плотности.

Равновесная плотность рассчитана при давлении 0.1 МПа для температур в диапазоне от 243.15 до 333.15 K с использованием рассматриваемых потенциалов. На рис. 2 представлены полученные значения вместе с экспериментальными данными [49].

**3.4. Оценки вязкости с использованием различных методов**

Расчёт коэффициентов вязкости производился с использованием трех различных методов в диапазоне температур от 273.15 до 333.15 K и давлении 0.1 МПа с использованием потенциала GAFF.

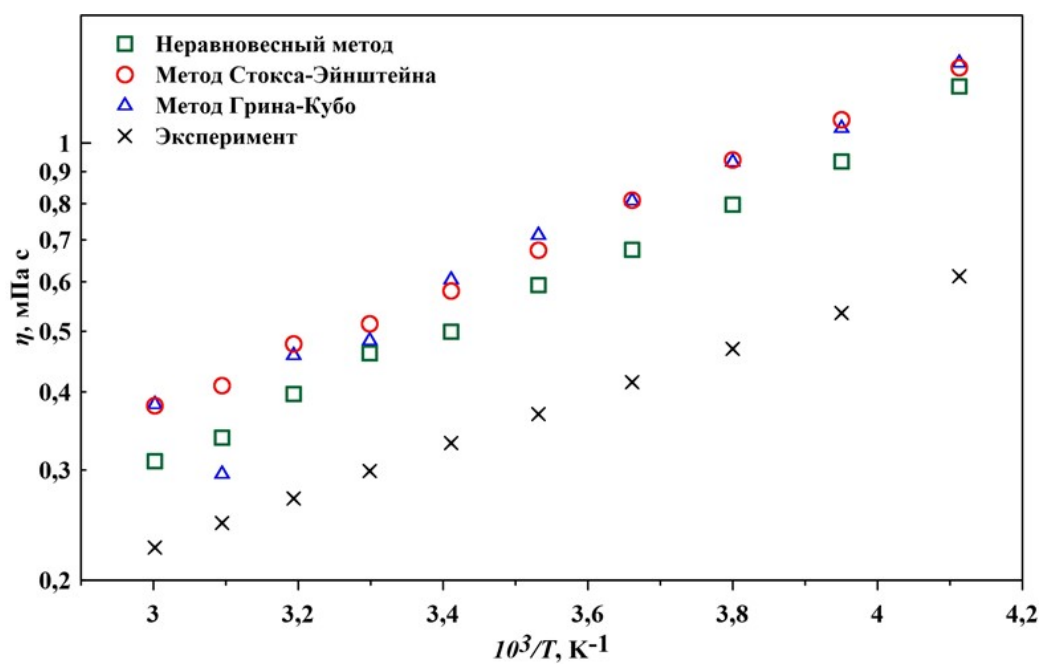

Рис. 3. Зависимость коэффициента вязкости от обратной температуры с использованием различных методов оценки вязкости. Крестики построены по экспериментальным данным из работы [49]. Сплошными линиями показаны аппроксимации зависимостей уравнением Аррениуса (1).

Зависимость коэффициента вязкости от температуры хорошо описывается уравнением Аррениуса:

$$\eta = \eta_0 \exp\left(\frac{E_a}{RT}\right), \qquad (1)$$

где $E_a$ – энергия активации, $R$ – универсальная газовая постоянная, $T$ – абсолютная температура системы, $\eta_0$ – предэкспоненциальный множитель. Соответственно логарифмическая зависимость вязкости от обратной температуры представляет собой линейную функцию. На рис. 3 представлены графики зависимости вязкости от обратной температуры, в логарифмическом масштабе.

### 3.5. Расчет коэффициентов вязкости с использованием различных потенциалов

Для тех же температур и давления, что и в предыдущем пункте, были рассчитаны коэффициенты вязкости с использованием потенциалов GAFF, OPLS-AA/CM1A и CHARMM36. Использовался метод Грина-Кубо, как наиболее распространенный и удобный для расчета вязкости. Результаты расчетов приведены на рис. 4.

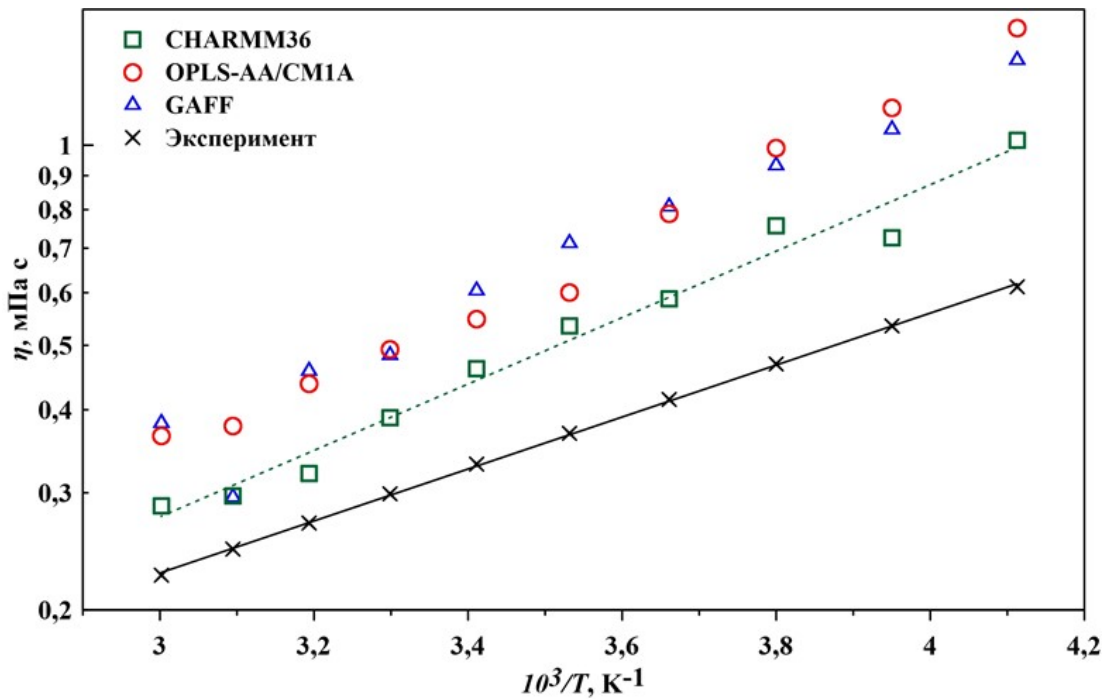

Рис. 4. Зависимость коэффициента вязкости от обратной температуры с использованием потенциалов AMBER, OPLS-AA и CHARMM. Крестики построены по экспериментальным данным из работы [49]. Прямыми линиями показаны аппроксимации зависимостей уравнением Аррениуса (уравнение (1)) для экспериментальных данных и CHARMM36, который обеспечивает наилучшие среди всех рассмотренных потенциалов значения вязкости.

# 4. ОБСУЖДЕНИЕ РЕЗУЛЬТАТОВ

Методом молекулярной динамики с использованием потенциалов GAFF, OPLS-AA/CM1A и CHARMM36 проведены расчёты некоторых свойств диизопропилового эфира для различных значений температуры и давления.

1. При оценке зависимости плотности диизопропилового эфира от давления (рис.1) полученные значения оказались несколько выше экспериментальных. Потенциалы GAFF и CHARMM36 дают схожие значения: для них отклонение не превышает 20 кг/м$^3$ (3 %). Для OPLS-AA отклонение не превышает 30 кг/м$^3$ (5%).

2. Оценка сжимаемости диизопропилового эфира при температуре 303.15 K (Таблица 1) показала неплохую сходимость с экспериментальными данными. При использовании потенциала CHARMM36 отклонение результатов молекулярного моделирования от эксперимента не превышает 12 %.

3. Расчет плотности диизопропилового эфира проводился в 10 точках в диапазоне температур от 243.15 до 333.15 K с использованием всех трех потенциалов. Результаты расчетов и экспериментальные данные приведены на рис. 2. Наилучшую точность расчета обеспечивает потенциалы GAFF и CHARMM36, которые дали близкие друг к другу значения. Для них отклонение от эксперимента не превышает 25 кг/м$^3$ (3.3 %). Наименьшую точность обеспечивает потенциал OPLS-AA/CM1A: ошибка составляет не более 40 кг/м$^3$ (5.1 %).

4. Расчёт коэффициента вязкости производился с использованием трех различных методов. Необходимость использования трех методов была вызвана тем, что первоначальные расчеты коэффициента вязкости диизопропилового эфира с использованием метода Грина-Кубо показали достаточно большое расхождение с экспериментальными данными. Поэтому возникла необходимость оценить этот параметр с использованием и других методов, чтобы убедиться, что ошибка вызвана недостаточной точностью математической модели, а не метода оценки вязкости. Результаты расчета вязкости с использованием 3 методов и данные эксперимента приведены на рис.3. Можно видеть, все три метода дали схожие значения. Это подтверждает, что ошибки оценки вязкости связаны не с ошибкой методов ее расчета, а с неточностью математической модели, вызванной не совсем точным соответствием используемых потенциалов реальным силовым полям, действующим внутри исследуемого вещества.

5. Сравнение результатов расчета вязкости с использованием трех потенциалов (рис. 4), показывает, что наибольшую точность моделирования вязкости обеспечивает потенциал CHARMM36. Значения вязкости, рассчитанные с использованием данного потенциала, отличаются от эксперимента на 20-70%, причем чем больше температура, тем выше точность. Ошибки оценки вязкости с использованием потенциалов GAFF и OPLS-AA/CM1A находятся в диапазоне 50-140 %. Таким образом можно рекомендовать использование потенциала CHARMM36 для исследования транспортных свойств диизопропилового эфира и его смесей с другими простыми эфирами если не требуется высокая точность расчетов.

## 5. ЗАКЛЮЧЕНИЕ

Потенциалы CHARMM36, GAFF и OPLS-AA/CM1A хорошо воспроизводят зависимость плотности и сжимаемости диизопропилового эфира от температуры и давления.

Среди исследованных потенциалов силовое поле CHARMM36 даёт наилучшее описание вязкости диизопропилового эфира и её зависимости от температуры. Это даёт основание предполагать, что данный потенциал должен давать удовлетворительное описание и других процессов переноса в диизопропиловом эфире, а также в смесях диизопропилового эфира с другими простыми эфирами среди которых наибольшей практический интерес в связи с задачей создания ион-селективных барьеров типа жидкой мембраны представляют растворы краун-эфиров в диизопропиловом эфире.

Тем не менее, нужно отметить, что все три потенциала дают заметное расхождение между экспериментальными значениями вязкости диизопропилового эфира и результатами молекулярно-динамического моделирования. Одной из причин этого расхождения может быть то, что параметры невалентных взаимодействий на атомах в органической молекуле универсальны для атома с каждым типом связи и не учитывают в полной мере индуктивного эффекта различных замещающих групп. Возможно, что при сохранении структуры рассмотренных потенциалов переопределение невалентных взаимодействий атомов с заданных в стандартной форме на рассчитанные индивидуально для исследуемой молекулы может помочь улучшить согласование между результатами молекулярно-динамического моделирования и реальным поведением вещества в эксперименте. Однако этот вопрос остаётся задачей дальнейшего исследования.

# 6. СПИСОК ЛИТЕРАТУРЫ